\documentclass{jps-cp}
\usepackage{txfonts} 

\title{The GlueX Experiment: Recent Results and Future Plans}

\author{Jonathan \textsc{Zarling}$^{1}$ \\
(For the GlueX Collaboration)
}

\inst{$^{1}$Department of Physics, Indiana University, Bloomington, IN 47401, USA}

\email{jzarling@indiana.edu}

\recdate{January 18, 2019}

\abst{
	The GlueX experiment, located in experimental Hall D of Jefferson Lab, seeks to map the spectrum of light mesons in search of exotic hybrid mesons, a predicted class of hadronic states with explicit gluonic degrees of freedom. GlueX is a photoproduction experiment utilizing a linearly polarized photon beam at $E_\gamma=$ 8-9 GeV and a large acceptance spectrometer. The experiment has already collected orders of magnitude more data than previous experiments at similar energies. We show results on asymmetry measurements in the production of pseudoscalars mesons, which refine our understanding of production mechanisms in photoproduction at our energies. We also present results on $J/\psi$ photoproduction, which provides model constraints on heavy quark photoproduction and the ability to search for s-channel production of pentaquark candidate $P_c^+$ states.
}
	
\kword{photoproduction, light meson, spectroscopy, asymmetry, GlueX, Jefferson Lab }

\begin{document}
\maketitle

\section{Introduction}

GlueX is a photoproduction experiment located at Thomas Jefferson Laboratory in Newport News, Virginia. Its primary goal is to further the understanding of light mesonic states and seek evidence for exotic mesons. Such states are characterized by quantum numbers $J^{PC}$ that cannot be explained using a quark-antiquark description, a clear indication of additional quark or valence gluonic components . Although many of these states may be expected from the underlying theory of quantum chromodynamics \cite{dudek}, experimental evidence for exotic $J^{PC}$ states in the light meson sector remains sparse \cite{meyer-swanson} . Utilizing wide acceptance and a high statistics data sample, states up to about 3 GeV/$c^2$ can be investigated using the GlueX spectrometer.

A major focus of present studies with the GlueX experiment is in understanding exchange mechanisms of photoproduction. Here, we focus on the production of single pseudoscalars and the $J/\psi$ meson to further our knowledge of production mechanisms at beam energies $E_\gamma = 8.5$ GeV.

%

\section{Experimental Apparatus}



The GlueX experiment utilizes a beam of photons produced through coherent Bremsstrahlung radiation with 11.6 GeV electrons from the Continuous Electron Beam Accelerator Facility (CEBAF) at Jefferson Laboratory on a diamond radiator. Figure 1 shows a schematic view of the GlueX spectrometer. Beam energy is determined from electron recoil momentum and photon polarization measured via triplet production ($\gamma e^- \rightarrow e^+e^-e^-$ in thin foil medium). This photon beam is incident on a 30 cm long liquid hydrogen target.  Charged particles produced are measured in forward and central drift chambers housed in a 2.2 T superconducting solenoid magnet. Neutral particles are measured in forward and barrel calorimeters. A time-of-flight counter provides particle identification, with a DIRC (Detection of Internally Reflected Cherenkov light) detector to be installed in early 2019, which will provide pion/kaon separation of at least 3$\sigma$ for momenta up to 4 GeV/c. Approximately 200 pb$^{-1}$ of data was collected as of December 2018. 

These components allow for the reconstruction of a wide range of hadronic final states with an acceptance of approximately 1-120$^\circ$ in laboratory $\theta$.

\begin{figure} \label{f1}
	\centering \includegraphics[scale=0.325]{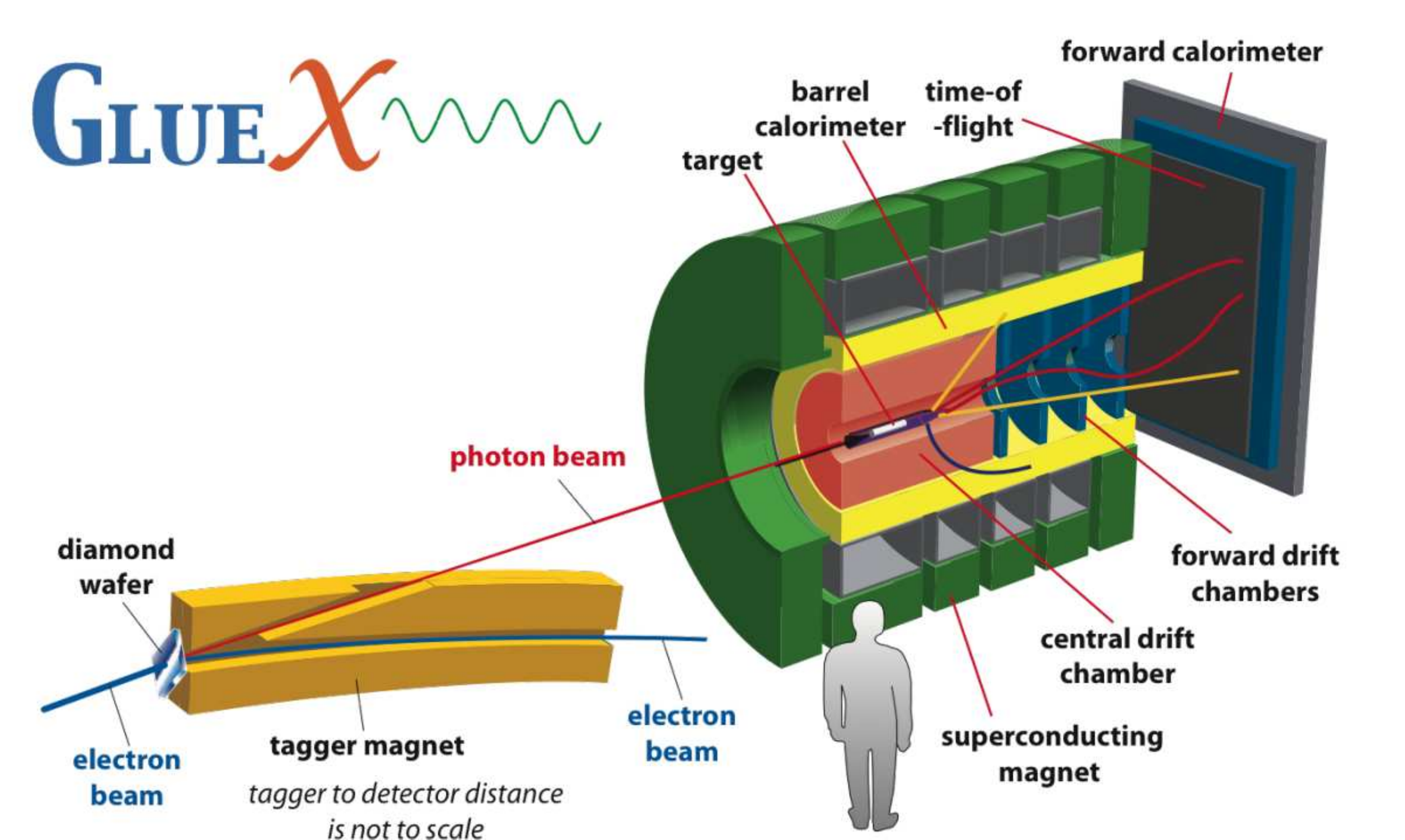}
	\caption{
		A sketch of the GlueX detector located at Jefferson Laboratory.
	}
\end{figure} 

\section{Early GlueX Analysis}


\subsection{Pseudoscalar Beam Asymmetries}

Beam asymmetries allow for an understanding of the exchange processes in photoproduction. At energies of 8.5 GeV, these exchanges can be modeled as $t$-channel processes involving the exchange of quasi-particle reggeons, which may vary as a function of momentum transfer variable $t$. For the case of producing a single pseudoscalar, the production can be fully encoded in a single physics observable denoted as $\Sigma$. With a linearly polarized photon beam, the photoproduction cross section can be expressed as $\sigma_{pol} = \sigma_{unpol} \left[ 1-P_\gamma\Sigma \mathrm{cos}(2\phi)  \right] $. Here, $P_\gamma$ is the beam polarization and $\sigma_{unpol}$ is an unpolarized cross section. The quantity $\Sigma$ is a measure of the \textit{naturality} of exchange reggeons (defined as $N\equiv P(-1)^J$). Measuring $\Sigma=+1$ would indicate pure natural exchanges in a pseudoscalar's production, whereas measuring $\Sigma=-1$ would indicate purely unnatural exchanges. The quantity may depend on Mandelstam $t$ and can be compared to theory model predictions.

Measurements of the production of single $\pi^0$, $\eta$, and $\eta^\prime$ pseudoscalars with recoil proton, as well as the production of a $\pi^-$ meson with recoiling $\Delta^{++}$ are given here. Results for $\pi^0$ and $\eta$ mesons are given in Figure 2. Results for $\eta^\prime$ and $\pi^- \Delta^{++}$ production are expected for publication in 2019; preliminary results are given in Figure 3	.

\begin{figure} \label{f2}
	\centering \includegraphics[scale=0.31]{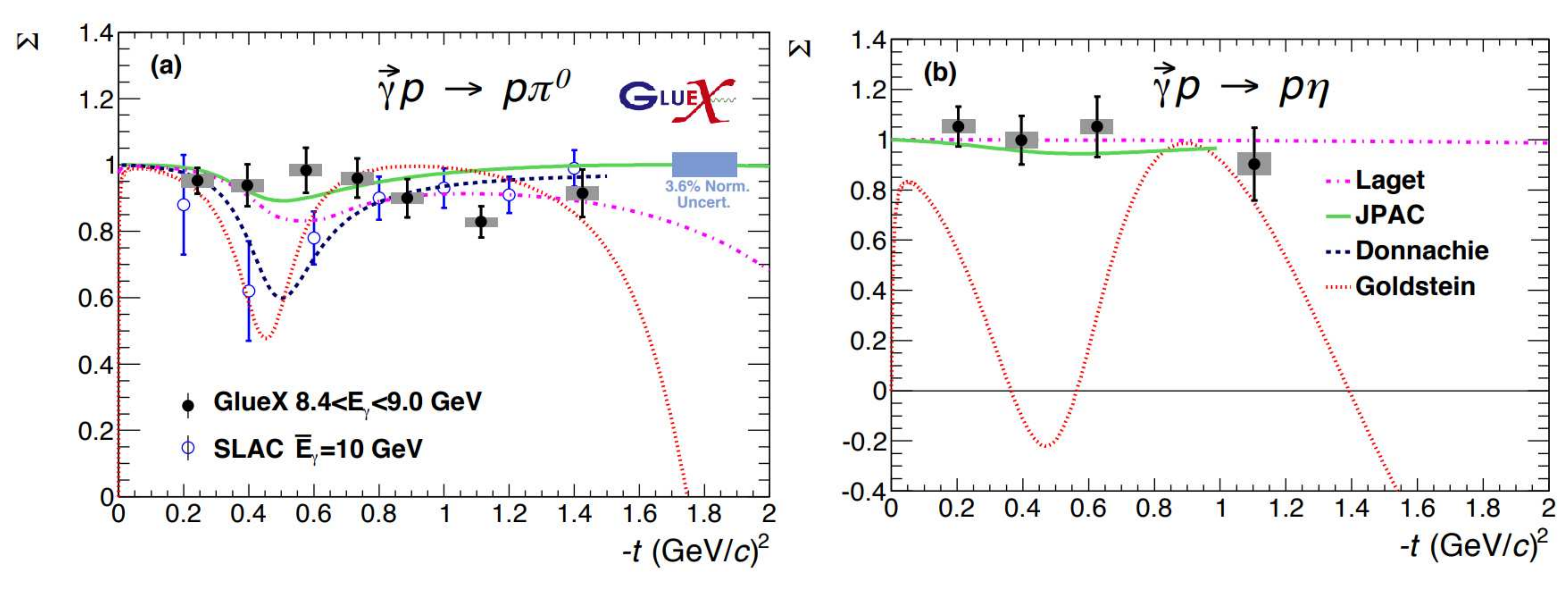}
	\caption{
		Asymmetry measurements for the production of $\pi^0$ and $\eta$ mesons decaying to two photons measured as a function of $t$. Colored curves indicate various theory models with more information available at \cite{pi0-paper}.
	}
\end{figure} 

\begin{figure} \label{f3}
	\centering \includegraphics[scale=0.55]{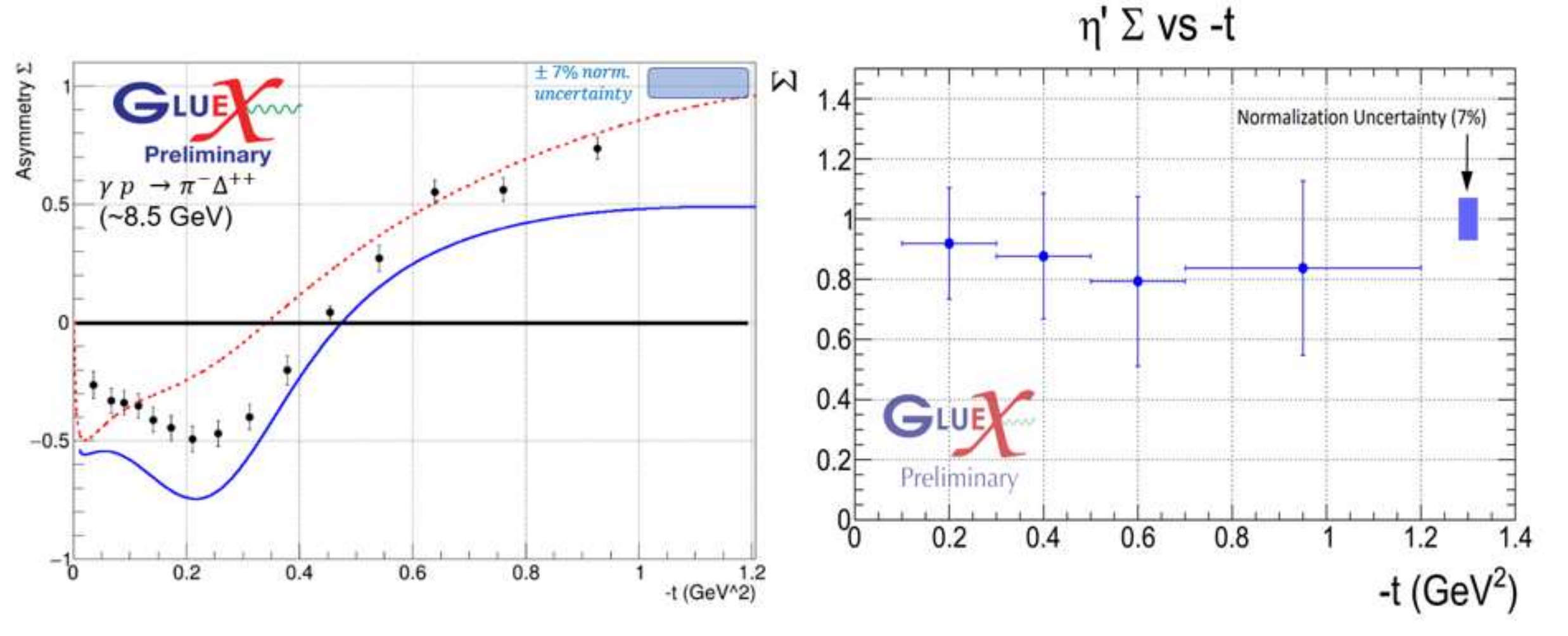}
	\caption{
		Asymmetry measurements for the production of $\pi^- \Delta^{++}$ and $\eta^\prime$ states. Colored curves on left plot indicate theory models from \cite{korea-th,jpac-piDelta} . Error bars shown indicate statistical uncertainties only.
	}
\end{figure} 

It is observed that in the neutral exchange processes studied, states are produced predominantly via natural exchange partners (e.g. $\rho$ and $\omega$ exchanges) with little significant $t$ dependence. In contrast, the charge exchange production of a $\pi^-\Delta^{++}$ system requires exchanges to have non-zero isospin and results demonstrate the contribution of both natural and unnatural exchanges with significant $t$ dependence. This indicates the exchange of unnatural reggeons (e.g. pion-like) predominantly at low $|t|$ and the exchange of predominantly natural exchanges (for instance $\rho$ and $a_2$ mesons) at higher $|t|$. These measurements will serve to inform future light meson searches with GlueX data. Additional work in studying the asymmetry of production for $a_0(980) p$, $K^+\Sigma^0$, and $K^+\Lambda(1520)$ states, along with cross sections for single pseudoscalar production will further improve the understanding of production mechanisms in the future.

%

\subsection{Threshold $J/\psi$ Production}

Photoproduction of the $J/\psi$ meson near threshold is also observed. Here previous data is very limited. Studies of the cross section as a function of incoming photon beam energy allows insight into the production of heavy quark states, which may be explained via two-gluon or three-gluon exchanges \cite{brodsky}. Our data will provide additional constraints to such models. Additionally, photoproduced $J/\psi$ mesons allow for direct searches for pentaquark candidate states reported by the LHCb experiment \cite{lhcb-pentaquark,lhcb-pentaquark2}, which may appear as an enhancement in the $J/\psi$ cross section at its threshold for production \cite{jpac-pentaquark}.

Figure 4 (left) shows the observed $e^+e^-$ spectra measured with the GlueX spectrometer. We see a clear peak indicating $J/\psi$ meson production near threshold energy. These are the first experimental results of the $J/\psi$ photoproduction cross section at threshold, as shown in Figure 4 (right). We do not observe clear indications of an enhancement in the cross section indicative of s-channel threshold production of narrow pentaquark states for $P_c^+(4312)$, $P_c^+(4440)$, or $P_c^+(4457)$ reported by the LHCb experiment with present statistics.

\begin{figure} \label{f4}
	\centering \includegraphics[scale=0.375]{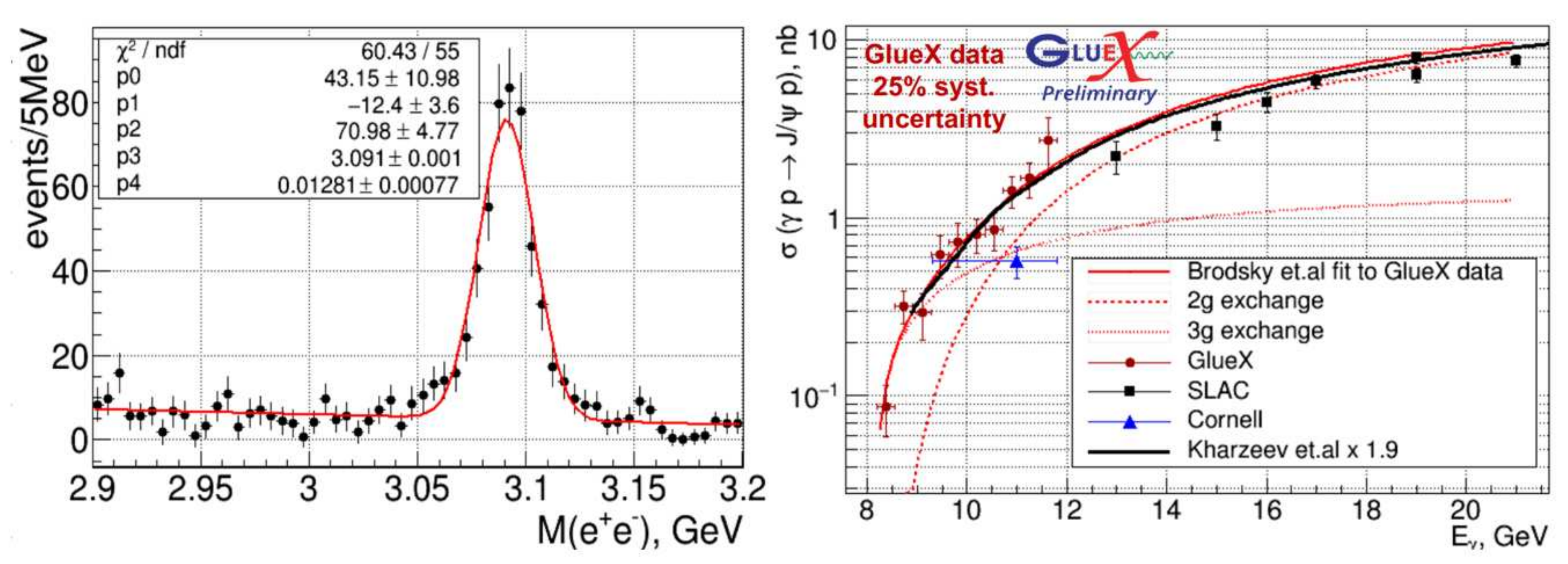}
	\caption{
		A plot of the $e^+e^-$ spectrum at GlueX in the narrow region around the $J/\psi$ mass (left) and preliminary results for the cross section of $J/\psi$ photoproduction as a function of energy (right). Measurements from SLAC and Cornell experiments and theoretical models are also included in the right plot.
	}
\end{figure}

%

\section{Conclusion}

GlueX is a unique experiment in the worldwide search for new states of hadronic matter, in large part due to its linearly polarized photon beam. We have summarized current results investigating the production mechanisms at play in photoproduction at GlueX energies. Pseudoscalar asymmetry measurements in both neutral and charge exchange reactions have begun to give an understanding of exchange mechanisms at play in photoproduction. These will be expanded in the future with asymmetry measurements in additional systems, analogous measurements of vector meson spin-density matrix elements, and cross section measurements. Additionally, we see a significant sample of $J/\psi$ events that enable us to study heavy quark threshold photoproduction and will allow us to observe or set limits on photocouplings for the reported $P_c^+$ pentaquark candidate states.

We have collected a few times more data than presented here and continue investigations in search of new and potentially exotic light meson states.

\section*{Acknowledgements}

We acknowledge the support of the US Department of Energy Office of Nuclear Physics under award DE-FG02-05ER41374. This material is based upon work supported by the U.S. Department of Energy, Office of Science, Office of Nuclear Physics under contract DE-AC05-06OR23177.

\end{document}